\documentclass[12pt,a4paper,aps,preprint,superscriptaddress,nofootinbib]{revtex4-1}
\usepackage[utf8]{inputenc}
\usepackage{graphicx}
\usepackage{amssymb}
\usepackage{textcomp}
\usepackage{amsmath}
\usepackage{tabularx}
\usepackage{bm}
\usepackage{times}
\usepackage{color}
\usepackage{slashed}
\usepackage{multirow}
\usepackage{verbatim}
\usepackage{cancel}
\usepackage{subfigure}
\usepackage{ulem}
\usepackage{lipsum,epsfig,dsfont}

\usepackage[colorlinks=true, pdfstartview=FitV, linkcolor=blue, citecolor=blue, urlcolor=blue]{hyperref}
\allowdisplaybreaks[4]

\def\lsim{\mathrel{\raise.3ex\hbox{$<$\kern-.75em\lower1ex\hbox{$\sim$}}}}
\def\gsim{\mathrel{\raise.3ex\hbox{$>$\kern-.75em\lower1ex\hbox{$\sim$}}}}

\newcommand{\be}{\begin{equation}}
	\newcommand{\ee}{\end{equation}}
\newcommand{\ba}{\begin{array}}
	\newcommand{\ea}{\end{array}}
\newcommand{\bea}{\begin{eqnarray}}
	\newcommand{\eea}{\end{eqnarray}}

\newcommand{\half}{\frac{1}{2}}

\newcommand{\eesasaa}{$e^+ e^- \to \slashed{\gamma}  \slashed{\gamma} \gamma$\,}
\newcommand{\eesesea}{$e^+ e^- \to \slashed{e}^+  \slashed{e}^- \gamma$\,}

\newcommand{\calC}{{\mathcal{C}}}

\newcommand{\MeV}{{\rm MeV}}

\definecolor{orange}{rgb}{1,0.5,0}

\usepackage[table]{xcolor}
\usepackage{colortbl}
\definecolor{mygray}{gray}{.95}

\begin{document}

	\title{Dark states with electromagnetic form factors at electron colliders}
	
	\author{Yu Zhang} 
	\affiliation{School of Physics, Hefei University of Technology, Hefei 230601,China}

	\author{Mao Song}
	\affiliation{School of Physics and Optoelectronics Engineering, Anhui University, Hefei 230601,China}
	
	\author{Liangwen Chen}
	\email{chenlw@impcas.ac.cn}
	\affiliation{Institute of Modern Physics, CAS, Lanzhou 730000, China}
	\affiliation{Advanced Energy Science and Technology Guangdong Laboratory, Huizhou 516000, China}
	
	\begin{abstract}

		Electromagnetically neutral dark sector particles may feebly interact with photons through higher dimensional effective operators, such as 
		mass-dimension 5 magnetic and electric dipole moment, and a mass-
		dimension 6 anapole moment and charge radius operators.
		In this work, we use hypercharge gauge field form factors to treat  dark states, which will induce not only electromagnetic form factors but also the corresponding $Z$ boson operators. Taking a Dirac fermion $\chi$ as an example,
		we investigate the probes of searching for such dark states at current and future $e^+e^-$ collider experiments including BESIII, STCF, Belle II and CEPC via monophoton searches. Comparing to current experiments, we find that electron colliders including BESIII, STCF, Belle II, which operate with the center-of-mass at several GeV, have leading sensitivity  on the corresponding electromagnetic form factors for the mass-dimension 5 operators with dark states lighter than several GeV,
		while they can not provide competitive upper limits  for the mass-dimension 6 operators.
		Future CEPC operated with the center-of-mass on and beyond the mass of $Z$-boson with competitive luminosity can probe the unexplored parameter space for dark states with mass-dimension 5 (6) operators in
		the mass region of $m\lesssim 100 $ GeV (10 MeV $\lesssim m\lesssim 100 $ GeV).

	\end{abstract}
	
	\maketitle
	
	\section{Introduction}
	\label{sec:Intro}
	
	Primary importance for our understanding of elementary interactions is shedding light on 
	the dark sector states.
	Searches for kinetically mixed dark photons and for new particles as  mediators  connecting to standard model (SM) constituent of the prime dark-sector physics cases \cite{Essig:2013lka}.
	However, some dark states sharing a coupling to the SM photon have received comparatively less attention.
	In this scenario, even if dark states are perfectly electromagnetic (EM) neutral, higher-dimensional effective couplings to the SM photons are still possible. 
	For dark states $\chi$ considered as  Dirac fermion, through various moments, such as magnetic dipole moments
	(MDM) and electric dipole moments (EDM) at mass-dimension 5, and anapole moment (AM) and charge radius interaction (CR) at mass-dimension 6, the couplings to the photon can be present \cite{Chu:2018qrm, Kavanagh:2018xeh}.
	
	In general, by interactions associated with photons, dark states $\chi$ with EM form factors
	can be produced, which therefore could be studied by accelerator-based experiments and stars.
	In Ref. \cite{Chu:2018qrm},  $\chi$ pair-production in electron beams on fixed targets at NA64 \cite{NA64:2017vtt}, LDMX \cite{LDMX:2018cma}, BDX \cite{BDX:2016akw}, and mQ \cite{Prinz:1998ua}
	has been studied.
	In addition, constraints from rare meson ($B$ and $K$) decays \cite{Bird:2004ts, E949:2004uaj, BNL-E949:2009dza}
	and SM precision observables such as $(g-2)_\mu$  
	and the running of the fine-structure constant are also worked out.
	The current limits on and detection prospects of 
	such dark sector particles $\chi$ 
	are illustrated in Ref. \cite{Chu:2020ysb}, utilizing high-intensity proton beams,  concretely from LSND \cite{LSND:1996jxj}, MiniBooNE-DM \cite{MiniBooNEDM:2018cxm}, CHARM-II \cite{CHARM-II:1989nic, CHARM-II:1994dzw}, and E613 \cite{Ball:1980ojt} and projected SHiP \cite{SHiP:2015vad} and DUNE \cite{DUNE:2015lol, DUNE:2017pqt}.
	Via their scattering off electrons in the Forward Liquid Argon Experiment detector at the LHC Forward Physics Facility, the prospects of searching for electromagnetically interacting $\chi$ particles have been studied in Ref. \cite{Kling:2022ykt}.
	A detailed astrophysical study of stellar cooling constraints for the mass of dark states $\chi$ dropping below MeV are further complemented in Ref. \cite{Chu:2019rok, Chang:2019xva}.
	Using missing energy searches at colliders, the constraints from $e^+e^-$ colliders BaBar \cite{Chu:2018qrm} and LEP \cite{Fortin:2011hv, Chu:2018qrm}, and 
	from proton-proton collisions at LHC are also investigated \cite{Fortin:2011hv, Arina:2020mxo}.
	
	In this work, we extend the analysis to other electron colliders operated at the GeV scale, including  BESIII \cite{Asner:2008nq}, Belle II \cite{Belle-II:2010dht}, and the proposed Super Tau Charm Factory (STCF) \cite{Luo:2019xqt, Charm-TauFactory:2013cnj,Shi:2020nrf}, which can probe the light dark states with the mass less than GeV scale. 
	In order to accurately measure the discovered Higgs, many high energy electron colliders are proposed, including Circular Electron Positron Collider (CEPC) \cite{CEPCStudyGroup:2018ghi},  Future Circular Collider $e^+e^-$ (FCC-ee) \cite{FCC:2018evy},  International Linear Collider (ILC) \cite{ILC:2019gyn}, and  Compact Linear $e^+e^-$ Collider (CLIC) \cite{Robson:2018enq}.
	We will take CEPC as an example to investigate the sensitivity on the dark state with EM form factor via higher-dimensional moments at future  high energy electron colliders.

	This paper is organized as follows. In Sec.~\ref{sec:model}, we introduce the interactions between the dark states and photons. In Sec.~\ref{sec:sb} we describe the signal and backgrounds of probing the dark state with electromagnetic form factor at electron colliders. In Sec.~\ref{sec:sllmZ} we present the constraints on the corresponding couplings at BESIII, STCF and Belle II that are operated with the center-of-mass (CM) energy $\sqrt{s}\ll M_Z$. In Sec.~\ref{sec:sgZ}, we consider the constraints  at CEPC that is operated with $\sqrt{s}\geq M_Z$. Our conclusions are drawn in Sec.~\ref{sec:con}.

	\section{Dark states with Electromagnetic form factor interactions}
	\label{sec:model}
	
	{In this work, we consider that the dark state $\chi$ is taken as a Dirac fermion, which  may have the effective interactions with the hypercharge gauge boson field $B_\mu$ as \cite{Chu:2018qrm,Arina:2020mxo},
		\bea
		{\cal L}_\chi=\frac{1}{2} \mu_{\chi}^B \bar{\chi} \sigma^{\mu \nu} \chi B_{\mu \nu}
		+\frac{i}{2} d_{\chi}^B \bar{\chi} \sigma^{\mu \nu} \gamma^{5} \chi B_{\mu \nu}
		-a_{\chi}^B \bar{\chi} \gamma^{\mu} \gamma^{5} \chi \partial^{\nu} B_{\mu \nu}
		+b_{\chi}^B \bar{\chi} \gamma^{\mu} \chi \partial^{\nu} B_{\mu \nu}.
		\label{eq:lagB}
		\eea 
Here, 	 $B_{\mu\nu}\equiv\partial_\mu B_\nu-\partial_\nu B_\mu$ is the  hypercharge gauge field strength, 
$\mu_\chi^B$ and $d_\chi^B$ are the dimensional coefficients of the mass-dimension 5 MDM
and EDM interactions, expressed in units of the Bohr magneton 
$\mu_B\equiv e/(2m_e)$ with $e$ being the electric charge and $m_e$ being the electron mass, and
$\sigma_{\mu\nu}\equiv i[\gamma^\mu,\gamma^\nu]/2$;
$a_\chi^B$ and $b_\chi^B$ are the dimensional coefficients of the mass-dimension 6 AM
and CR interactions. 
Hypercharge form factors are linear combinations of electromagnetic form factors and the corresponding $Z$ boson operators, weighted by appropriate factors of the cosine and sine of the Weinberg angle $c_W$ and $s_W$.
Then Eq. (\ref{eq:lagB}) can be written with electromagnetic field
strength tensor $F_{\mu\nu}\equiv\partial_\mu A_\nu-\partial_\nu A_\mu$ and $Z$ gauge field strength tensor $Z_{\mu\nu}\equiv\partial_\mu Z_\nu-\partial_\nu Z_\mu$  as 
\bea
{\cal L}_\chi&=&\frac{1}{2} \mu_{\chi}^\gamma \bar{\chi} \sigma^{\mu \nu} \chi F_{\mu \nu}
+\frac{i}{2} d_{\chi}^\gamma \bar{\chi} \sigma^{\mu \nu} \gamma^{5} \chi F_{\mu \nu}
-a_{\chi}^\gamma \bar{\chi} \gamma^{\mu} \gamma^{5} \chi \partial^{\nu} F_{\mu \nu}
+b_{\chi}^\gamma \bar{\chi} \gamma^{\mu} \chi \partial^{\nu} F_{\mu \nu} \\ \nonumber
&+&\frac{1}{2} \mu_{\chi}^Z \bar{\chi} \sigma^{\mu \nu} \chi Z_{\mu \nu}
+\frac{i}{2} d_{\chi}^Z \bar{\chi} \sigma^{\mu \nu} \gamma^{5} \chi Z_{\mu \nu}
-a_{\chi}^Z \bar{\chi} \gamma^{\mu} \gamma^{5} \chi \partial^{\nu} Z_{\mu \nu}
+b_{\chi}^Z \bar{\chi} \gamma^{\mu} \chi \partial^{\nu} Z_{\mu \nu},
\label{eq:lagaZ}
\eea 
with $\calC_\chi^\gamma=\calC_\chi^B c_W$ and $\calC_\chi^Z=-\calC_\chi^B s_W$ where $\calC_\chi=\mu_\chi,\, d_\chi,\, a_\chi,\, b_\chi$.

In the scenarios of energies far below the electroweak scale, the $Z$
boson degree of freedom decouples and the effective interactions of Eq. (\ref{eq:lagB}) or (\ref{eq:lagaZ}) can be identically induced to
	\bea
{\cal L}_\chi=\frac{1}{2} \mu_{\chi} \bar{\chi} \sigma^{\mu \nu} \chi F_{\mu \nu}
+\frac{i}{2} d_{\chi} \bar{\chi} \sigma^{\mu \nu} \gamma^{5} \chi F_{\mu \nu}
-a_{\chi} \bar{\chi} \gamma^{\mu} \gamma^{5} \chi \partial^{\nu} F_{\mu \nu}
+b_{\chi} \bar{\chi} \gamma^{\mu} \chi \partial^{\nu} F_{\mu \nu},
\label{eq:lag}
\eea
which have been investigated in Refs.  \cite{Chu:2018qrm, Kavanagh:2018xeh,Chu:2020ysb,Kling:2022ykt,Chu:2019rok, Chang:2019xva,Fortin:2011hv}.
The usual electromagnetic form factors should be denoted by the $\gamma$ superscript,
which will be omitted in the following for simplicity unless otherwise stated.
}

	\section{Signal and Background at $e^+e^-$ colliders}
	\label{sec:sb}
	
	The cross section for the single photon 
	production from $e^+e^-$ annihilation, $e^+ e^- \to \chi \bar\chi \gamma$,
	can be approximately factorized into the process without
	photon emission, 
	$e^+ e^- \to \gamma/Z\to \chi \bar\chi$,
	times the improved 
	Altarelli-Parisi radiator function 
	\cite{Nicrosini:1989pn,Montagna:1995wp},
	\begin{equation}
		\frac{d^2\sigma}{d x_\gamma d z_\gamma}=
		H\left(x_\gamma, z_{\gamma}; s\right) \sigma_{0}(s_\gamma)\,,
		\label{eq:diffxs}
	\end{equation}
	where the radiator function is
	\begin{equation}
		H\left(x_\gamma, z_{\gamma} ; s\right)=\frac{\alpha}{\pi}\frac{1}{x_\gamma} \left[\frac{1+(1-x_\gamma)^2}{1-z_{\gamma}^{2}}- \frac{x_\gamma^2}{2}\right]\,.
	\end{equation} 
	Here, ${s}$ and ${s_\gamma}$ are the square of the center-of-mass (CM) energies of the $e^+e^-$ and $\chi\bar\chi$ system, respectively, with $s_\gamma=(1-x_\gamma)s$, $E_\gamma$ is the energy of the initial state
	radiation (ISR) photon,  $x_\gamma=2E_\gamma/\sqrt{s}$ is the energy fraction emitted away by ISR, $z_\gamma=\cos\theta_\gamma$ with $\theta_\gamma$ being the polar angle of the photon. The cross section of the $\chi$ pair production without ISR $\sigma_{0}$ reads
	\bea
	\label{eq:sig0}
	\sigma_{0}(s)&=&\frac{\alpha}{4} \frac{f\left(s\right)}{s^2}\sqrt{\frac{s-4 m_{\chi}^{2}}{s}}\Bigg[ c_W^2+ (g_L+g_R)\frac{s(s-M_Z^2)}{(s-M_Z^2)^2+M_Z^2\Gamma_Z^2}  \nonumber \\ 
	&+&\half\frac{1}{c_W^2}(g_L^2+g_R^2)\frac{s^2}{(s-M_Z^2)^2+M_Z^2\Gamma_Z^2}\Bigg],
	\eea
	with $g_L=-\half+s_W^2$ and $g_R=s_W^2$, and $M_Z$ and $\Gamma_Z$ are the mass and decay width of the $Z$-boson.
	The factor $f(s)$ is given as
	\bea
	\mathrm{MDM}: f\left(s\right)=\frac{2}{3} (\mu_{\chi}^{B})^2 s^{2}\left(1+\frac{8 m_{\chi}^{2}}{s}\right), \\
	\mathrm{EDM}: f\left(s\right)=\frac{2}{3} (d_{\chi}^{B})^2 s^{2}\left(1-\frac{4 m_{\chi}^{2}}{s}\right), \\
	\mathrm{AM}: f\left(s\right)=\frac{4}{3} (a_{\chi}^{B})^2 s^{3}\left(1-\frac{4 m_{\chi}^{2}}{s}\right), \\
	\mathrm{CR}: f\left(s\right)=\frac{4}{3} (b_{\chi}^{B})^2 s^{3}\left(1+\frac{2 m_{\chi}^{2}}{s}\right).
	\eea
{In the following, we will present the results of  usual electromagnetic form factors $\calC_\chi$ with $\calC_\chi=\calC_\chi^B c_W$.}
	
	It can be found that, in the limit of $\frac{m_\chi^2}{s}\to 0$,  the production rates
	of the $\chi$ pair with the EM form factors for mass-dimension-5 MDM and EDM operators have the same forms. So do for mass-dimension-6 AM and CR operators. When $\sqrt{s}\ll M_Z$, the production rate for $\chi$ pair
	in Eq.~(\ref{eq:sig0}) tends to be same with the one only considering dark sector-photon interactions \cite{Chu:2018qrm}.

	For monophoton searches at electron colliders, the backgrounds consist of two categories:  the irreducible background and the reducible background.  The irreducible background arises from the neutrino pair production associated with one visible photon $e^+e^-\to\nu\bar\nu\gamma$.  The reducible background comes from one visible photon in the final state together with several other SM particles that cannot be detected because of the detector limitations. The reducible background will be discussed later in details for each experiment, since it strongly depends on the detector performance.

	\section{$e^+e^-$ colliders operated with $\sqrt{s}\ll M_Z$}
	\label{sec:sllmZ}

	\subsection{Belle \uppercase\expandafter{\romannumeral2}}\label{sec:belle2}

	The constraints on the light dark states with electromagnetic form factors from Belle II via monophoton searches have been investigated in Ref.~\cite{Chu:2018qrm}, in which the authors follow Ref.~\cite{Essig:2013vha}, and scale up the BaBar background from Ref.~\cite{BaBar:2008aby} to an integrated luminosity of
	50 ab$^{-1}$ with the CM energy of 10.57 GeV, employ a
	constant efficiency cut of 50\% in both search regions and
	take identical geometric cuts for Belle II and BABAR.
	Ref. \cite{Kou:2018nap} has provided the exact background subtraction in the monophoton search of Belle II to probe an  
	invisibly decaying dark photon.
	In this work, we revisit the constraints from Belle II via monophoton searches following the strategy in Ref. \cite{Kou:2018nap}.
	
	At Belle II, the Electromagnetic Calorimeter (ECL), which covers a polar angle region of $(12.4-155.1)^{\circ}$
	and has inefficient gaps between the barrel and the endcaps  for polar angles
	between $(31.4-32.2)^{\circ}$ and $(128.7-130.7)^{\circ}$ in the lab frame \cite{Kou:2018nap}.
	The photons and electrons can be detected in the ECL.
	In the monophoton signature, the reducible background at Belle II consists of  
	two major parts \cite{Liang:2019zkb}:
	one is mainly due to the lack of polar angle  coverage 
	of the ECL near the beam direction, i.e., $\theta>155.1^{\circ}$ or $\theta<12.4^{\circ}$, which is referred to 
	as the ``bBG''; 
	the other one is mainly  owing to the gaps between 
	the three segments in the ECL detector, i.e., $\theta\in(31.4-32.2)^{\circ}$ or $(128.7-130.7)^{\circ}$,  
	which is referred to as the ``gBG''.

	The bBG  comes from the electromagnetic processes $e^+e^-\to \gamma +\slashed{X}$,
	dominated by $e^+e^-\to\gamma\slashed{\gamma}(\slashed{\gamma})$ and $e^+e^-\to\gamma\slashed{e}^+\slashed{e}^-$.
	Here, $\slashed{X}$ denotes the other particle(s) in the final state are emitted along the beam directions.
	Thus, except the single detected photon all the other final state particles 
	are emitted along the beam directions with 
	$\theta>155.1^{\circ}$ or $\theta<12.4^{\circ}$ in the lab frame, which are out of  the cover polar angle region of the ECL. 
	
	At the asymmetric Belle II detector, for the monophoton events from reducible bBG,
	the maximum energy of the final photon  
	in the CM frame $E_\gamma^m$,  is given by \cite{Liang:2019zkb, Zhang:2020fiu}
	(if not exceeding $\sqrt{s}/2$) 
	\begin{equation}
		E_\gamma^m(\theta_\gamma) = 
		\frac{ \sqrt{s}(A\cos\theta_1-\sin\theta_1)}
		{A(\cos\theta_1-\cos\theta_\gamma)-(\sin\theta_\gamma+\sin\theta_1)},
		\label{eq:bBG}
	\end{equation}
	where  
	all angles are given in the CM frame, 
	and $A=(\sin\theta_1-\sin\theta_2)/(\cos\theta_1-\cos\theta_2)$, 
	with $\theta_1$ and $\theta_2$ being 
	the polar angles corresponding to 
	the edges of the ECL detector.
	To remove the nasty bBG, the detector cut 
	\begin{equation}
		E_\gamma^{\mathrm{CM}} > E_\gamma^m  
	\end{equation}
	is adopted for the final monophoton (hereafter the {\it``bBG cut"}),
	with $E_\gamma^{\mathrm{CM}}$ being the photon energy in the CM frame.
	{Noting that, the ``bBG" in the reducible background can be eliminated 100\% by the  {\it``bBG cut"} theoretically without considering other possible backgrounds  that are caused by	instruments. }

	In the gBG, the monophoton energy can be quite large around  
	$\theta_{\gamma}\sim 0$ region, because the gaps in the  ECL are significantly away from the beam direction.
	The simulations for gBG have been carried out by 
	Ref. \cite{Kou:2018nap} to search for an  
	invisibly decaying dark photon.
	Two different sets of detector cuts
	are designed to optimize the detection efficiency for different masses of
	the dark photon:
	the {\it``low-mass cut"} and {\it``high-mass cut"}.
	The {\it``low-mass cut"} can be described as $\theta_{\rm min}^{\rm low}
	< \theta_{\gamma}^{\rm lab}<\theta_{\rm max}^{\rm low}$, where 
	$\theta_{\rm min}^{\rm low}$ and $\theta_{\rm max}^{\rm low}$
	are the minimum and maximum angles for the photon in the lab frame, respectively
	fitted as functions of \cite{Duerr:2019dmv}
	\begin{eqnarray}
		\theta_{\mathrm{min}}^{\mathrm{low}} &=& 5.399^{\circ} (E_\gamma^{\mathrm{CM}})^{2} / \mathrm{GeV}^{2}-58.82^{\circ} E_\gamma^{\mathrm{CM}} / \mathrm{GeV}+195.71^{\circ}, \\
		\theta_{\mathrm{max}}^{\mathrm{low}}  &=& -7.982^{\circ} (E_\gamma^{\mathrm{CM}})^{2} / \mathrm{GeV}^{2}+87.77^{\circ} E_\gamma^{\mathrm{CM}} / \mathrm{GeV}-120.6^{\circ}.
	\end{eqnarray}
	The {\it``high-mass cut"} can be described as $\theta_{\rm min}^{\rm high}
	< \theta_{\gamma}^{\rm high}<\theta_{\rm max}^{\rm high}$, where 
	$\theta_{\rm min}^{\rm high}$ and $\theta_{\rm max}^{\rm high}$
	can be respectively
	fitted as functions of \cite{Duerr:2019dmv}
	\begin{eqnarray}
		\theta_{\mathrm{min}}^{\mathrm{high}} &=& 3.3133^{\circ} (E_\gamma^{\mathrm{CM}})^{2} / \mathrm{GeV}^{2}-33.58^{\circ} E_\gamma^{\mathrm{CM}} / \mathrm{GeV}+108.79^{\circ}, \\
		\theta_{\mathrm{max}}^{\mathrm{high}}  &=& -5.9133^{\circ} (E_\gamma^{\mathrm{CM}})^{2} / \mathrm{GeV}^{2}+54.119^{\circ} E_\gamma^{\mathrm{CM}} / \mathrm{GeV}-13.781^{\circ}.
	\end{eqnarray}

	In order to probe the sensitivity for the light dark states with electromagnetic form factors
	at Belle II, we use the definition 
	{
	\bea
	\chi^2(\calC_\chi)\equiv  \frac{S^2(\calC_\chi)}{S(\calC_\chi)+B+(\epsilon B)^2},
	\eea
	where $S$ ($B$) is the number of events in the signal (background) processes, $\epsilon$ is the background systematic uncertainty}, and $\calC_\chi = \mu_\chi,\, d_\chi,\, a_\chi,\, b_\chi$ denotes the dimensional coefficient 
	of MDM, EDM, AM and CR interactions, respectively. For background, $B=B_{\rm ir}+B_{\rm re}$ consists of the  number of events 
	in irreducible background $B_{\rm ir}$ and reducible background $B_{\rm re}$.
	The  number of events $B_{\rm ir}$ ($S$) can be obtained from the  irreducible background (signal) 
	by integrating the differential cross section in the phase space regions under the related detector cuts,
	and assuming photon detection efficiency as 95\% \cite{Kou:2018nap}.
	It is found that about 300 (25000) gBG events 
	survived the  {\it``low-mass cut"} ({\it``high-mass cut"}) with 
	20 fb$^{-1}$ integrated luminosity  \cite{Kou:2018nap} in the reducible background, which will be rescaled according to the considered luminosity. 
{We show the numbers for an ideal case with zero systematics ($\epsilon=0$) and also for a possible case with 10\% systematics ($\epsilon=10\%$).}

\begin{figure*}[htbp]
		\begin{centering}
			\includegraphics[width=0.45\columnwidth]{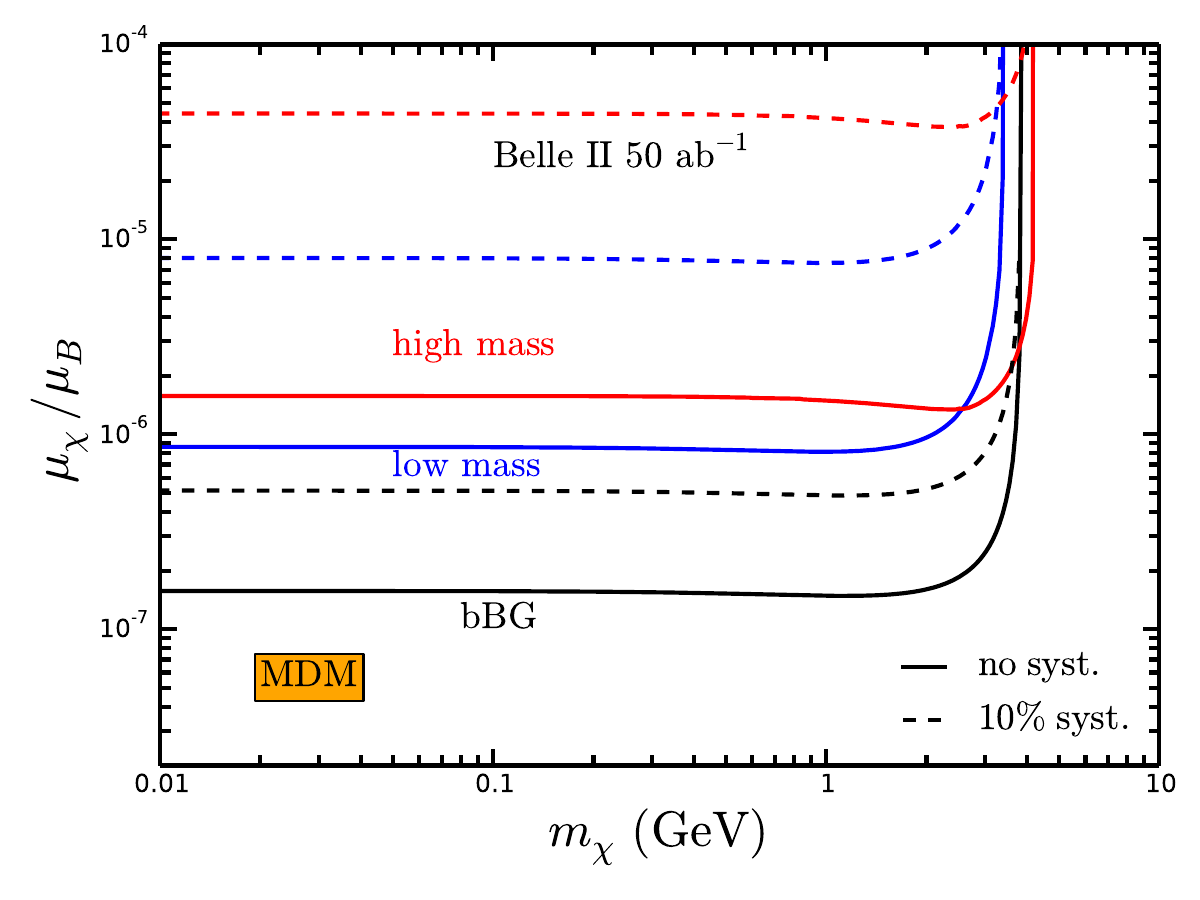}
			\includegraphics[width=0.45\columnwidth]{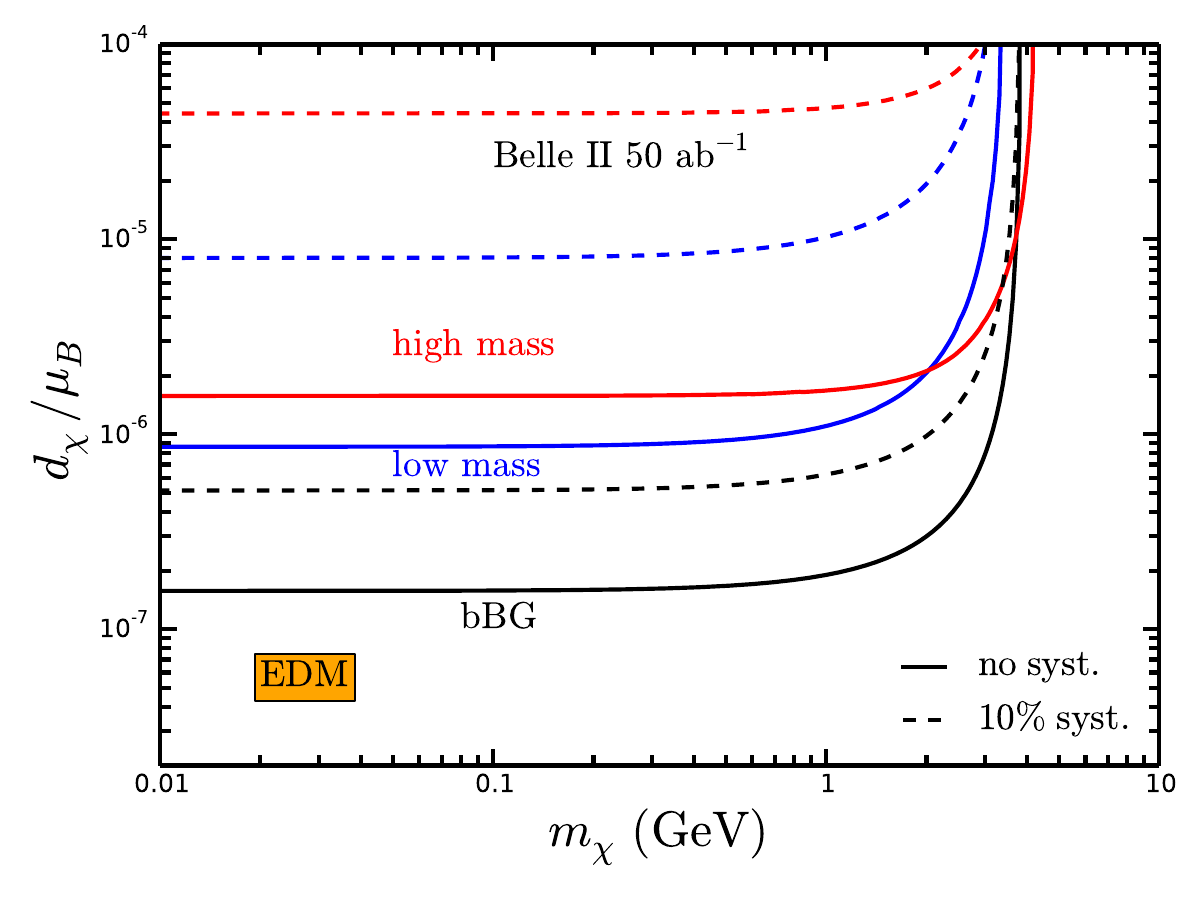}			
			\includegraphics[width=0.45\columnwidth]{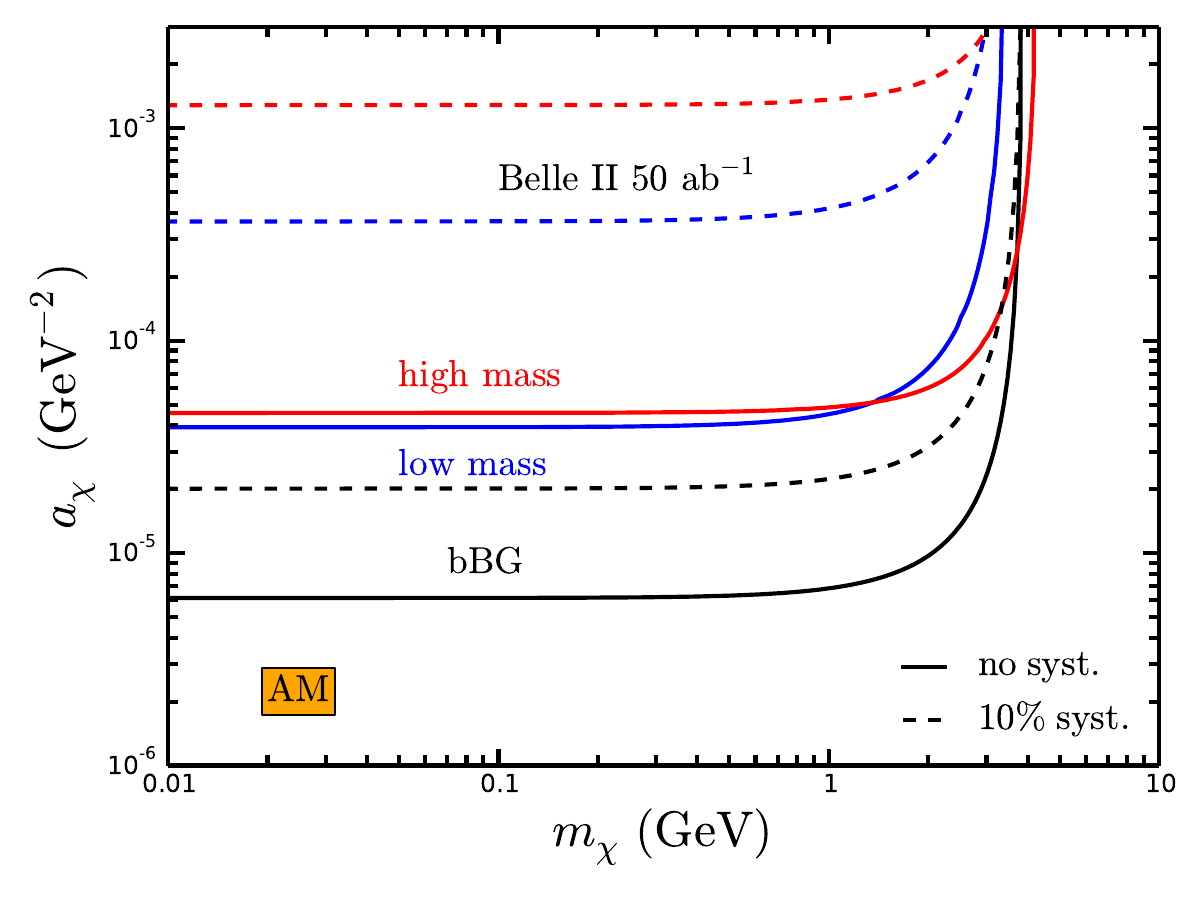}
			\includegraphics[width=0.45\columnwidth]{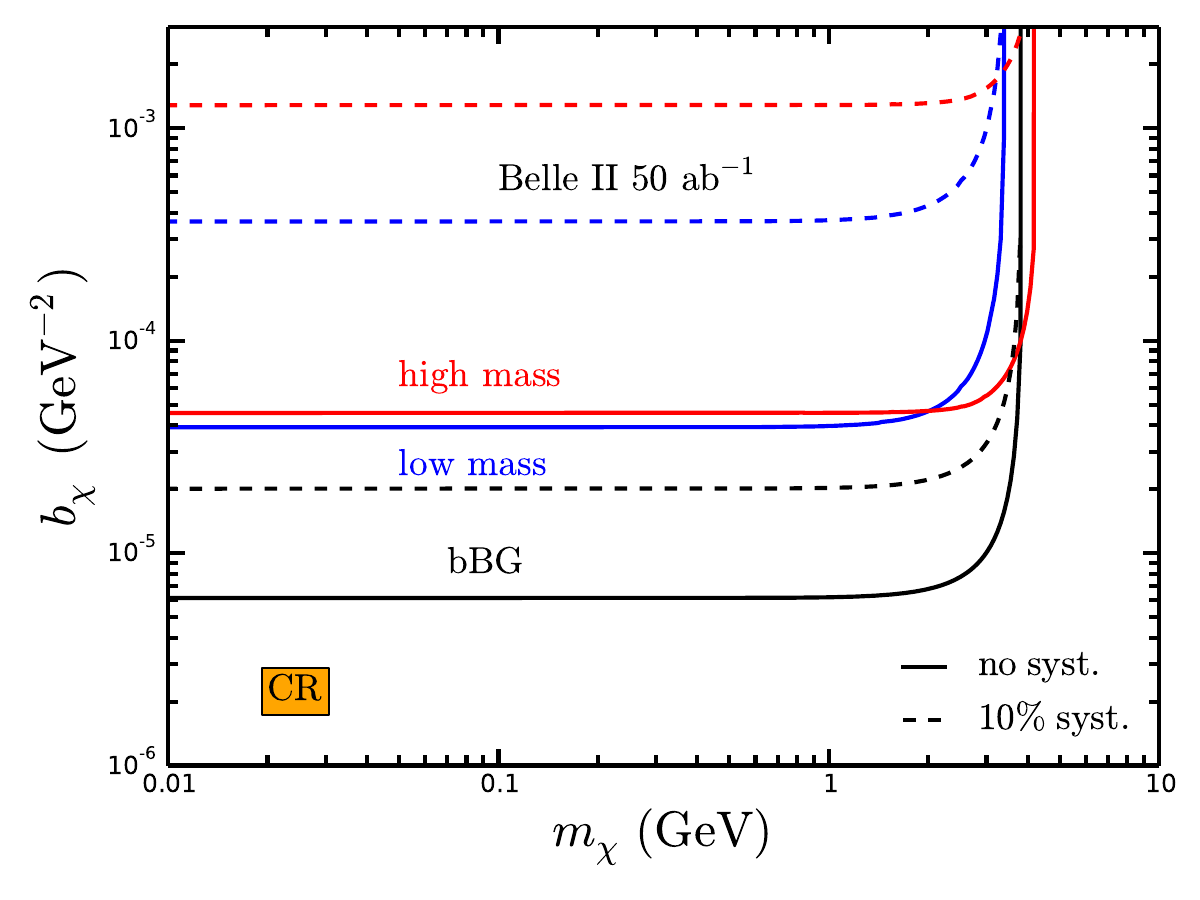}
			\caption{			
				The expected 95\% C.L. upper limits on the electromagnetic form factors for  mass-dimension 5 operators  through MDM (top left) and EDM (top right), and  mass-dimension 6 operators  through AM (bottom left) and CR interaction (bottom right), at Belle II under {\it``low-mass cut"} (blue), {\it``high-mass cut"} (red) and
				{\it``bBG cut"} (black), with 50 ab$^{-1}$ integrated luminosity. The solid (dashed) lines are assuming zero (10\%) background systematics.}
			\label{fig:belle2}
		\end{centering}
	\end{figure*}

	By solving $\chi^2(\calC_\chi)-\chi^2(0)=2.71$, one can obtain the 95\% confidence level (C.L.) 
	upper limits on the corresponding electromagnetic form factors for dark states.
	The upper limits under  {\it``low-mass cut"} and {\it``high-mass cut"} at Belle II with 50 ab$^{-1}$ integrated luminosity  are shown in Fig. \ref{fig:belle2}.
	We can see that, {assuming zero background systematics} the constraints under  {\it``low-mass cut"} are better than  {\it``high-mass cut"} 
	for MDM with $m_\chi\lesssim$ 2.6 GeV, EDM with $m_\chi\lesssim$ 2.1 GeV, AM with $m_\chi\lesssim$ 1.4 GeV,
	and CR with $m_\chi\lesssim$ 2.0 GeV.
	The upper limits of the relevant interaction coefficients at 50 ab$^{-1}$ Belle II can down to about $8.6\times 10^{-7}\, \mu_B$ ($8.0\times 10^{-6}\, \mu_B$), and $3.9\times 10^{-5}$ GeV$^{-2}$ ($3.6\times 10^{-4}$ GeV$^{-2}$) for light dark state with mass-dimension 5 and mass-dimension 6 operators {with zero (10\%) background systematics}, respectively. {It is seen that for an ideal case with zero systematics, the sensitivity of Belle II on the electromagnetic form factors for dark states can improve about one order relative to the case with 10\% systematics. Thus the control on the systematic uncertainty on the background  is very important.}

	In order to investigate the effects of the gBG, and compare with other experiments where  detailed simulations with gBG being not available, the 95\% C.L. upper limits without taking gBG into account are also presented in Fig. \ref{fig:belle2} {labeled as bBG}. In this scenario, the reducible background can vanish with the {\it``bBG cut"}, and now the background events are all
	provided by the irreducible backgrounds that survived the {\it``bBG cut"}. We find that the upper bound under  {\it``bBG cut"} can be down to about $1.6\times 10^{-7}\, \mu_B$ 
	($6.1\times 10^{-6}$ GeV$^{-2}$) for mass-dimension 5 (6) operators, which is about five (six) times stronger {than} the one when gBG is considered under  {\it``low-mass cut"}.
	{It should be noted that the the Belle II bBG results are not to be interpreted
		as any actual sensitivity possible by current Belle II experiment, while can explore the potential sensitivity of Belle II, if  a new subdetector can detect the particles emitting from the
		gaps in ECL.}
	
	\subsection{BESIII and STCF}
	
	The proposed STCF~\cite{Luo:2019xqt} in China is a symmetric double ring electron-positron collider. It is the next generation tau charm facility and  successor  of  the BESIII  experiment,
	and designed to have CM energy ranging from 2 to 7 GeV.
	At BESIII and STCF, the cut for the final photon: 
	$E_\gamma > 25\ \MeV$ in the barrel ($|z_\gamma|<0.8$)  or 
	$E_\gamma > 50\ \MeV$ in the end-caps ($0.86<|z_\gamma|<0.92$) \cite{Ablikim:2017ixv}
	is applied.
	Besides, we use the BESIII detector parameters to analyze the projected
	constraints from STCF due to the similarity of these two experiments.
	Since there is no released analysis  on gBG at BESIII  so far as we know, gBG in the monophoton reducible background at BESIII and STCF is not considered. 
	Without taking gBG into account,  
	the monophoton reducible background at  BESIII and STCF, mainly arises from \eesasaa and \eesesea. At symmetric BESIII and STCF, we also apply the  detector cut \cite{Liu:2019ogn, Zhang:2019wnz}:
	\begin{equation}
		E_\gamma >E_\gamma^m(\theta_\gamma)= \frac{\sqrt{s}}{(1+{\sin\theta_\gamma}/{\sin\theta_b})},
		\label{eq:adv-cuts}
	\end{equation}
	on the final state photon to remove the reducible background,
	where  
	$\theta_b$ denotes
	the angle at the boundary of the sub-detectors.
	Taking into account the coverage of main drift
	chamber, electromagnetic calorimeter, and
	time-of-flight, we have the polar angel $\cos{\theta_b}=0.95$ at BESIII \cite{Liu:2018jdi}.
	The photon detection efficiency is assumed as $100\%$ at both BESIII and STCF in this work,
	since photon reconstruction
	efficiencies are all more than 99\% \cite{BESIII:2011ysp} at BESIII.
	
\begin{figure*}[htbp]
		\begin{centering}
			\includegraphics[width=0.45\columnwidth]{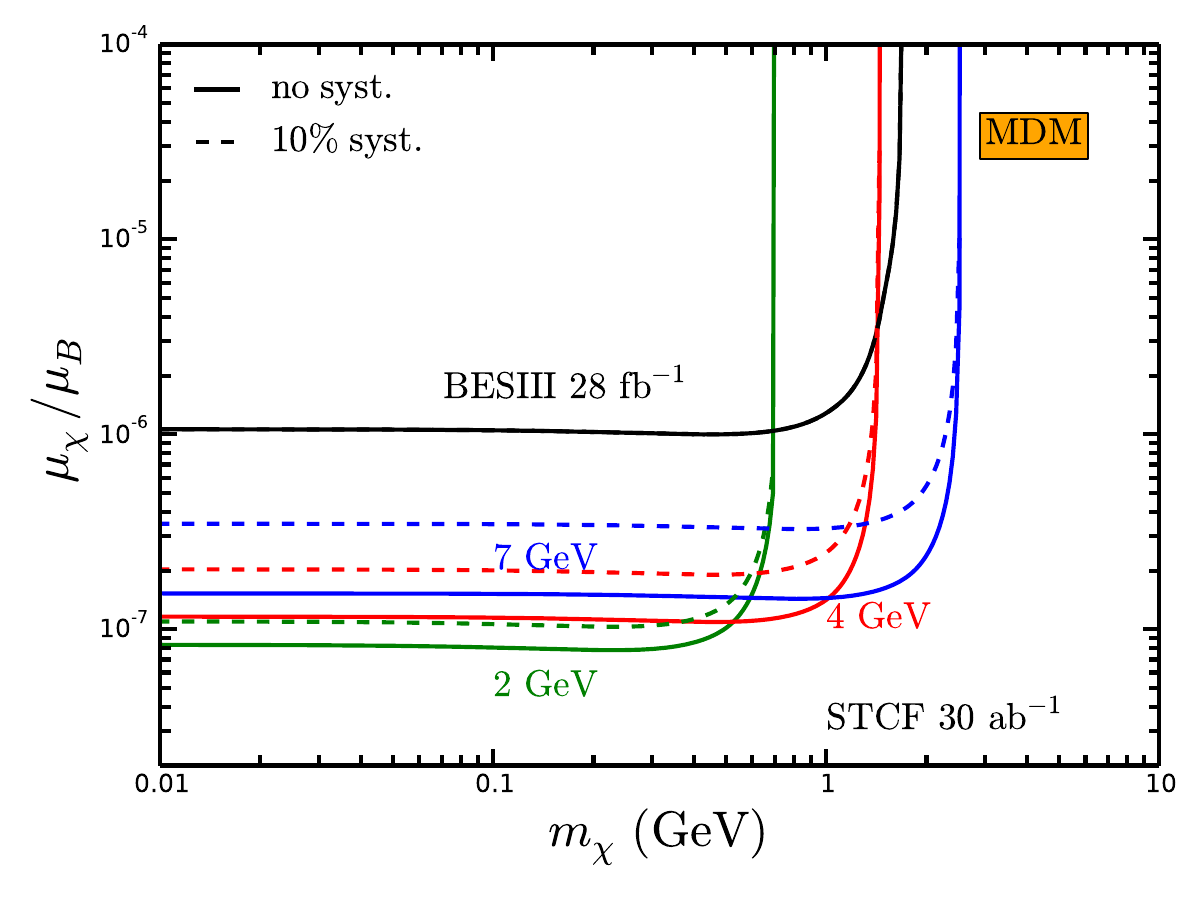}
			\includegraphics[width=0.45\columnwidth]{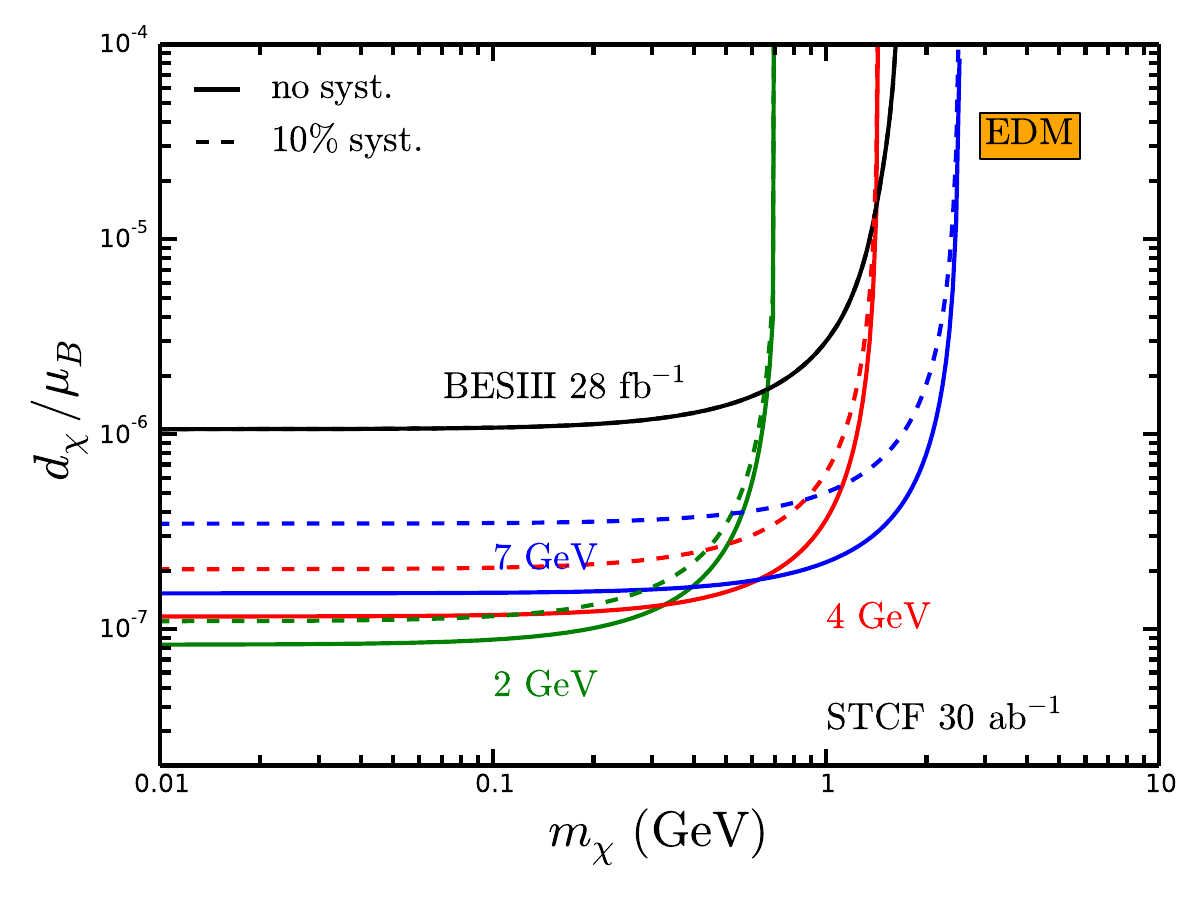}			
			\includegraphics[width=0.45\columnwidth]{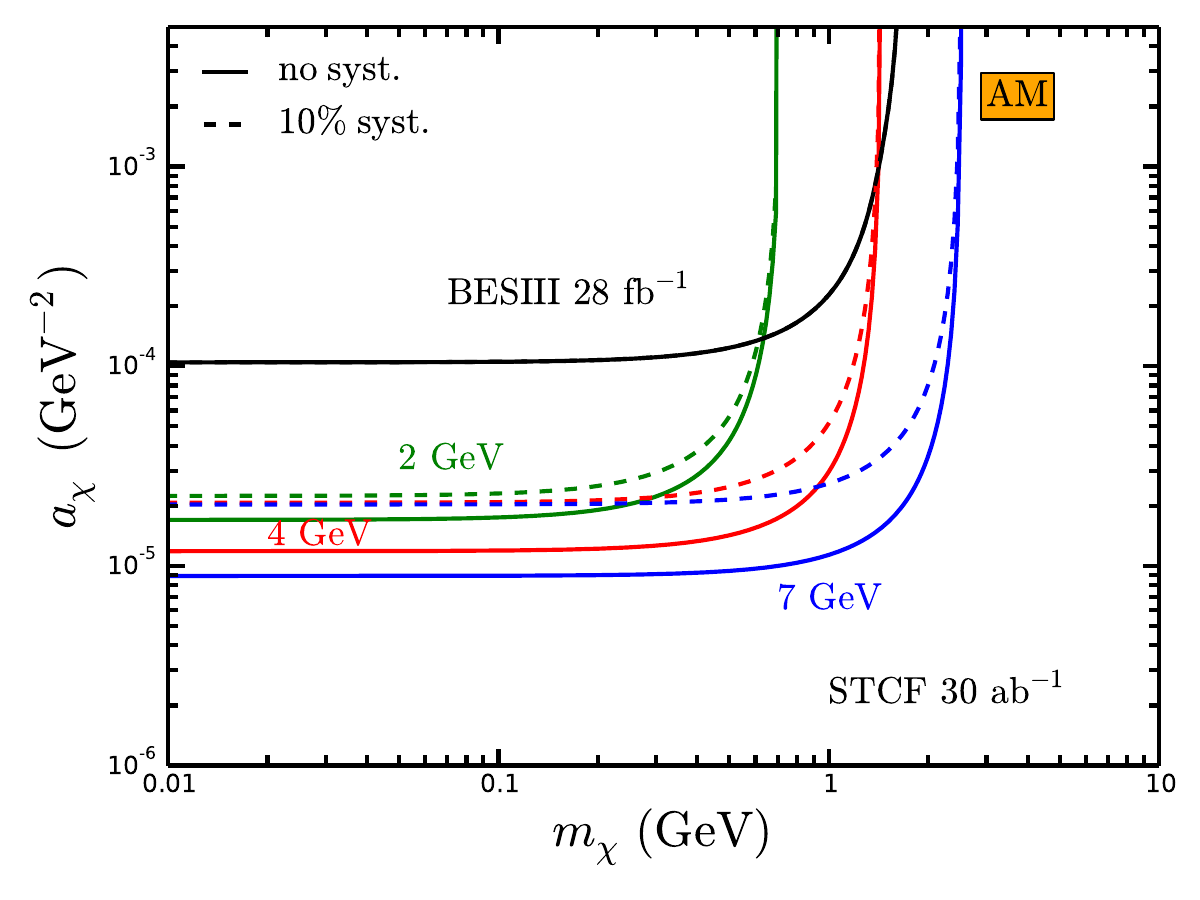}
			\includegraphics[width=0.45\columnwidth]{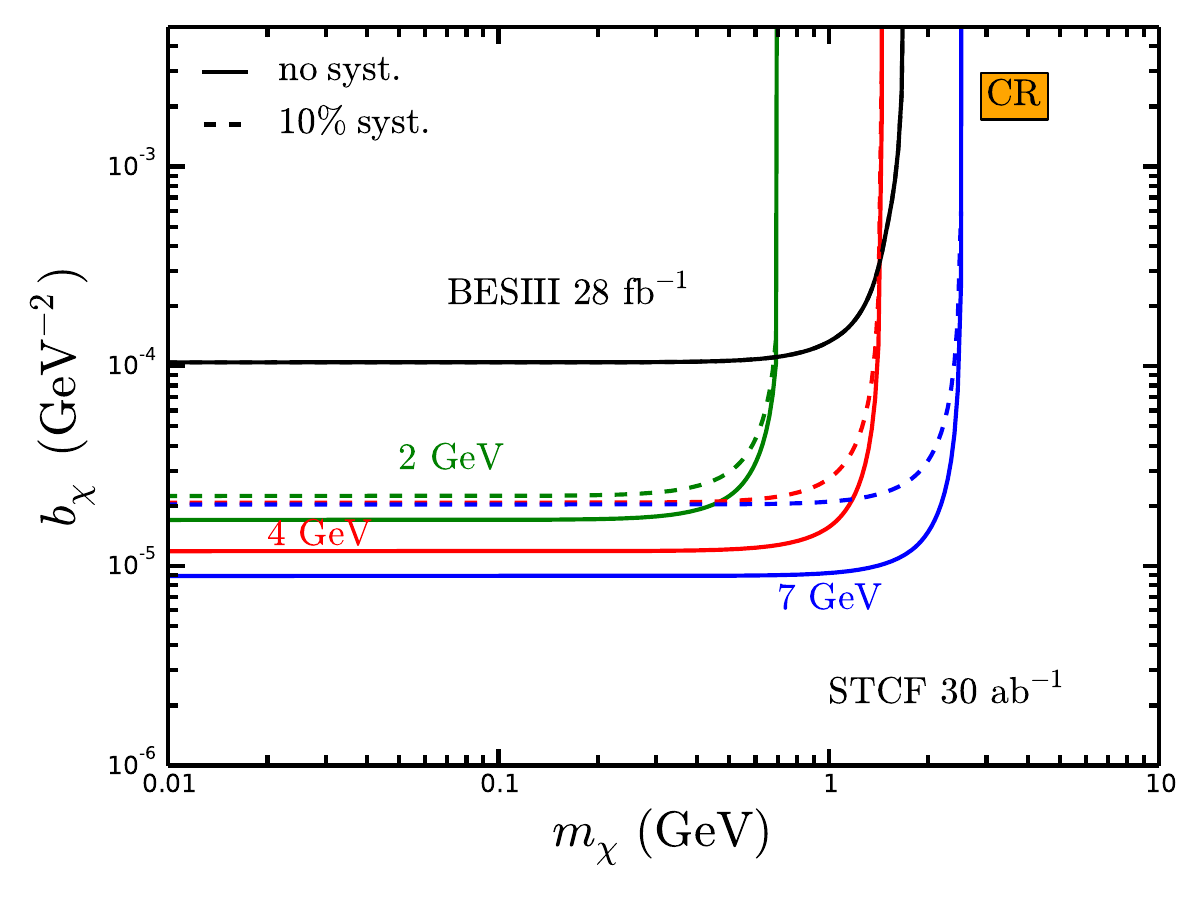}	
			\caption{
				The expected 95\% C.L. upper limits on the electromagnetic form factors at BESIII and STCF, 			
				for mass-dimension 5 operators (left) though MDM (top left) and EDM (top right), and  mass-dimension 6 operators  through AM (bottom left) and CR interaction (bottom right).
				The limits from BESIII (black) is obtained with about 28 fb$^{-1}$ integrated luminosity collected at the various CM energies ranging
				from 2.125 GeV to 4.95 GeV during 2012-2021. The expected limits from STCF are shown at three typical energy points with 30 ab$^{-1}$ integrated luminosity for
				$\sqrt{s}$= 2 (green), 4 (red), and 7 GeV (blue), respectively.
			The solid (dashed) lines are assuming zero (10\%) background systematics}
			\label{fig:bes3}
		\end{centering}
	\end{figure*}

	Since 2012, the monophoton trigger has been
	implemented at BESIII, and the corresponding events have been collected with the luminosity of about  28 fb$^{-1}$ at the CM energy ranging from 2.125 GeV to 4.95 GeV  until 2021.
	We compute the number of events due to signal ($S$) and backgrounds ($B$) under the 
	applied cuts, and define $\chi_{\rm tot}^2(\calC_\chi)=\sum_i\chi_{i}^2(\calC_\chi)$,
	where $\chi_{i}^2(\calC_\chi)\equiv S_i^2/(S_i+B_i+(\epsilon B_i)^2)$ with index $i$ denoting each BESIII colliding energy.
	The expected  95\% C.L. upper limits 
	on  the electromagnetic form factors of the light dark fermion $\chi$ according to about 28 fb$^{-1}$ luminosity
	collected at BESIII are shown in Fig. \ref{fig:bes3} by demanding $\chi_{\rm tot}^2(\calC_\chi)= \chi^2(0)+2.71$.
	Fig. \ref{fig:bes3} also shows the expected 95\% C.L. upper limits  with assumed 30 ab$^{-1}$ luminosity 
	at three typical colliding energies, $\sqrt{s}= 2, 4, 7$ GeV, in future STCF, respectively. {The  solid (dashed) lines are assuming zero (10\%) background systematics.}
	We find that, BESIII can probe couplings down to about $1.1\times 10^{-6}\, \mu_B$  for light dark states  with mass-dimension 5 operators and down to $1.0\times 10^{-4}$ GeV$^{-2}$ with mass-dimension 6 operators.
	{The assuming 10\% background systematics do not affect much on the results at BESIII, since the backgrounds mainly  from the irreducible background with gBG omitted are not significant.}
	With the same luminosity, operated at lower
	energy, STCF has better sensitivity  in probing the light dark fermion $\chi$ with the electromagnetic form factors though mass-dimension 5 operators.
	This is because the monophoton cross section in small mass $\chi$ production is not very dependent on the CM energy, while  in the background decreases with the increment of the CM energy.
	On the contrary, for mass-dimension 6 operators, the production rates of light dark states are even more sensitive to the
	center-of-mass energy, thus higher energy STCF has better sensitivity.
	
	{It should be noted that, as far as we know, BESIII has not released any analysis
		on the gBG. Thus  for the BESIII and
	STCF analyses, the gBG  is temporarily neglected in this work. 
	Improved BESIII and STCF limits can
	be obtained in the future when the gBG analysis is
	available.}

	\section{$e^+e^-$ colliders operated with $\sqrt{s}\geq M_Z$}
	\label{sec:sgZ}

	\subsection{LEP}
	
	The monophoton searches have been investigated carefully by all four LEP experiments \cite{Workman:2022ynf}.
	In this work,we consider the limits on the cross section presented by L3 Collaboration, both
	on $Z$-pole \cite{L3:1998uub} with an integrated luminosity of 100 pb$^{-1}$ at CM energies $\sqrt s=89.45-91.34$ GeV, and off $Z$-pole  with 619 pb$^{-1}$ at $\sqrt s=188.6-209.2$ GeV \cite{L3:2003yon}. 
	Using the L3 off Z-pole data for the monophoton searches  \cite{L3:2003yon}, the bounds 
	in the presence of the $\chi$ couplings to only photon via  mass-dimension 5 MDM and EDM operators, and  mass-dimension 6 AM and CR operators have been studied in Refs.~\cite{Fortin:2011hv, Chu:2018qrm}.
	In this work, we revisit the sensitivity at LEP on the $\chi$ couplings to not only photon but also $Z$ boson via Eq. (\ref{eq:lagB}) using the L3 measurements both on \cite{L3:1998uub} and off \cite{L3:2003yon} $Z$-pole.
	For comparison of cross sections at $Z$-pole, we require photon 
	energy within the range 1 GeV $<E_\gamma<$ 10 GeV and the angular acceptance $45^{\circ}<\theta_\gamma<135^{\circ}$, following the 
	same event acceptance criteria as in Ref. \cite{L3:1998uub} with six data subsets.
	Similarly, for the off $Z$-pole analysis, 
	the high-energy photon should  lie in the kinematic region $14^{\circ}<\theta_\gamma<166^{\circ}$, and $p_T^\gamma>0.02\sqrt{s}$,
	following the same event topology as described in Ref. \cite{L3:2003yon} with eight data subsets.
	We obtain the bounds on the couplings  from 
	the data subset that leads to the best constraints using $|\sigma^{\rm SM}+\sigma^\chi-\sigma^{\rm exp}|\leq \delta\sigma^{\rm exp}$.
	Here $\sigma^{\rm SM}$ is the SM cross section, $\sigma^\chi$ represents the contribution from $\chi\bar\chi\gamma$ production, and $\sigma^{\rm exp}\pm \delta\sigma^{\rm exp}$ denotes the experiment result.
	We find that the off $Z$-pole measurement imposes more
	stringent bound than the $Z$-pole measurement does, which can be seen in Fig. \ref{fig:cepc}.

	Due to the couplings with $Z$ boson via Eq. (\ref{eq:lagB}),  the dark sectors with $m_\chi < M_Z/2$ can now be constrained by invisible $Z$ decay. The $Z$ boson partial decay widths into $\chi\bar\chi$ mediated by hypercharge form factors can be expressed as 
	\bea
	\Gamma_{Z\to\chi\bar\chi}=\frac{s_W^2 f(M_Z^2)}{16\pi M_Z}\sqrt{\frac{M_Z^2-4 m_{\chi}^{2}}{M_Z^2}}.
	\eea
	The total width of the $Z$ boson has been measured accurately by the LEP experiments which place a strong bound on beyond-the-SM contributions $
	\Gamma_{Z\to\chi\bar\chi}< 2.0$ MeV at 95\% C.L. \cite{ALEPH:2005ab}. 
	In Fig. \ref{fig:cepc}, we show the constraints on the electromagnetic form factors for $\chi$ from the measurement  of invisible $Z$ decay at LEP. 
	It can be found that  the limits on dark states with electromagnetic form factors by invisible $Z$ decay are stricter than those by monophton searches at LEP for MDM with $m_\chi\lesssim$ 45 GeV, EDM with $m_\chi\lesssim$ 40 GeV, AM with $m_\chi\lesssim$ 25 GeV, and CR with $m_\chi\lesssim$ 40 GeV.

	\subsection{CEPC}

	In the following, we will focus on the  future CEPC~\cite{CEPCStudyGroup:2018rmc, CEPCStudyGroup:2018ghi}.
	The CEPC, proposed by the Chinese high
	energy physics community in 2012, is designed to run primarily at a center-of-mass energy
	of 240 GeV as a Higgs factory ($H$-mode) with a total  luminosity of $ 20\ \mathrm{ab}^{-1}$ for ten years running. 
	In addition, it will also be operated on the $Z$-pole as a $Z$
	factory ($Z$-mode) with a total luminosity of $100\ \mathrm{ab}^{-1}$ for two years running at $\sqrt{s}=91.2\ \mathrm{GeV}$, perform a precise $WW$ threshold scan ($WW$-mode) with a total luminosity of $\sim 6\ \mathrm{ab}^{-1}$ for one year running at 
	$\sqrt{s} \sim$ $160\ \mathrm{GeV}$, and be upgraded to a center-of-mass energy
	of 360 GeV, close to the $t\bar t$ threshold ($t\bar t$-mode) with a total luminosity of $\sim 1\ \mathrm{ab}^{-1}$ for five years \cite{CEPCPhysicsStudyGroup:2022uwl}. 
	
	{The monophoton signature,  where the large missing transverse momentum carried away by the $\chi\bar\chi$ pair is balanced by a final state visible photon, is used  to probe the dark states. Following the CEPC CDR~\cite{CEPCStudyGroup:2018ghi}, the visible photon need  to satisfy the cuts $|z_\gamma|<0.99$ and $E_\gamma > 0.1 $  GeV. 
 Beyond the irreducible background from  the neutrino pair production in association with a visible photon $e^+e^-\to\nu\bar\nu\gamma$, any SM process with a single photon in the final state can contribute to the total background, with all other visible particles  undetected. Since the SM processes which contain either jets or charged particles are relatively easy to distinguish from a dark state event, their contribution to the total background is negligible \cite{Bartels:2012ex, Habermehl:2020njb}. 
 However, the exception is for the  radiative Bhabha scattering, $e^+e^-\to e^+e^-\gamma $, which has a huge cross section and can mimic the signal if both the final state electrons and positrons escape undetected, for example, through the beam pipes. In our following analysis, we consider both neutrino and radiative Bhabha backgrounds. }

	{To remove the monophoton events in the reducible background from radiative Bhabha process $e^+e^-\to e^+e^-\gamma$, 	we apply the cut 
	\begin{equation}
	E_\gamma >E_\gamma^m(\theta_\gamma)= \frac{\sqrt{s}}{(1+{\sin\theta_\gamma}/{\sin\theta_b})},
	\label{eq:adv-cuts}
	\end{equation}
	on the final state photon following Ref. \cite{Liu:2019ogn}, where  $\theta_b$ corresponds to
	the polar angle at the boundary of the sub-detectors with $\cos\theta_b=0.99$.	For certain polar angle $\theta_\gamma$, the maximum energy of the final photon $E_\gamma^m$ in the reducible background occurs when the final state electron and positron emit along different beam directions with $\theta_{e^\pm}=\theta_b$.
}

\begin{figure*}[htbp]
	\begin{centering}
		\includegraphics[width=0.45\columnwidth]{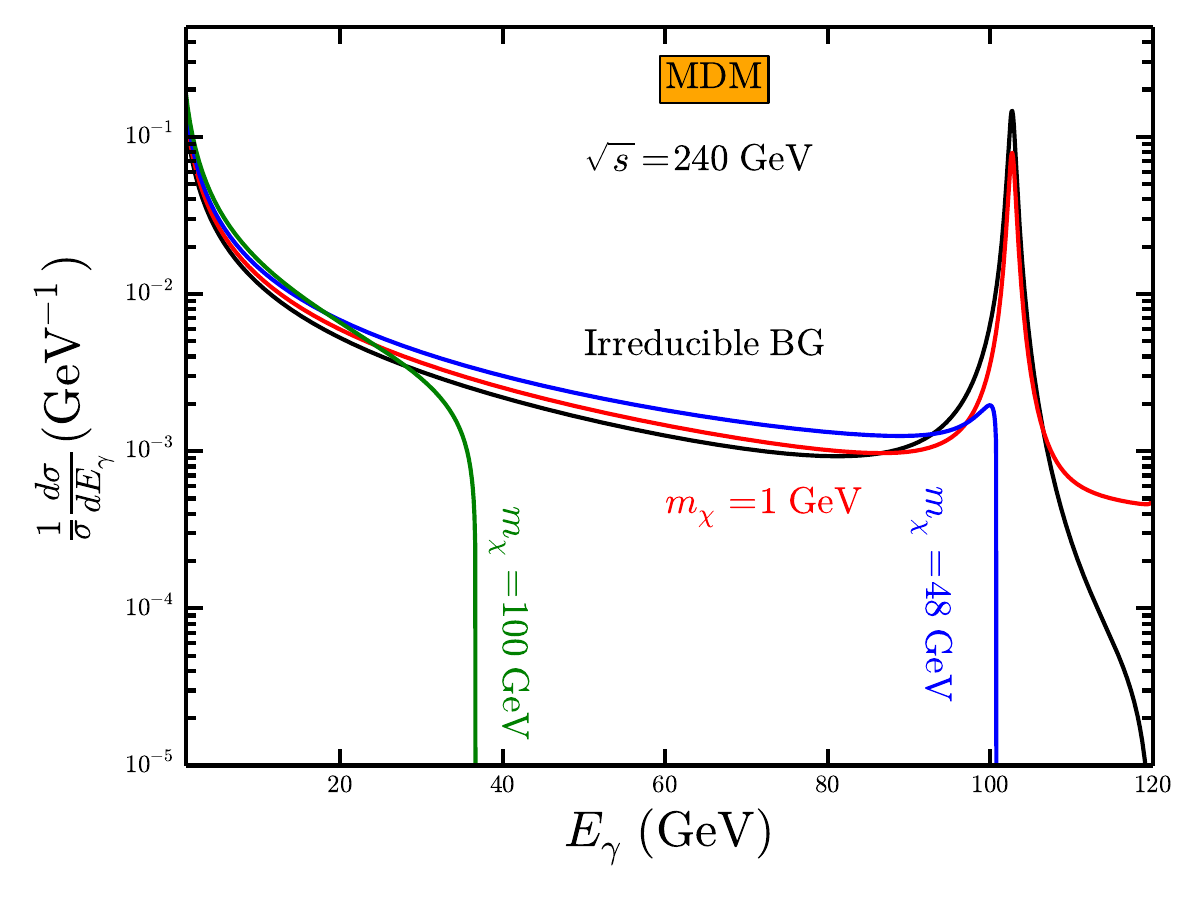}
		\includegraphics[width=0.45\columnwidth]{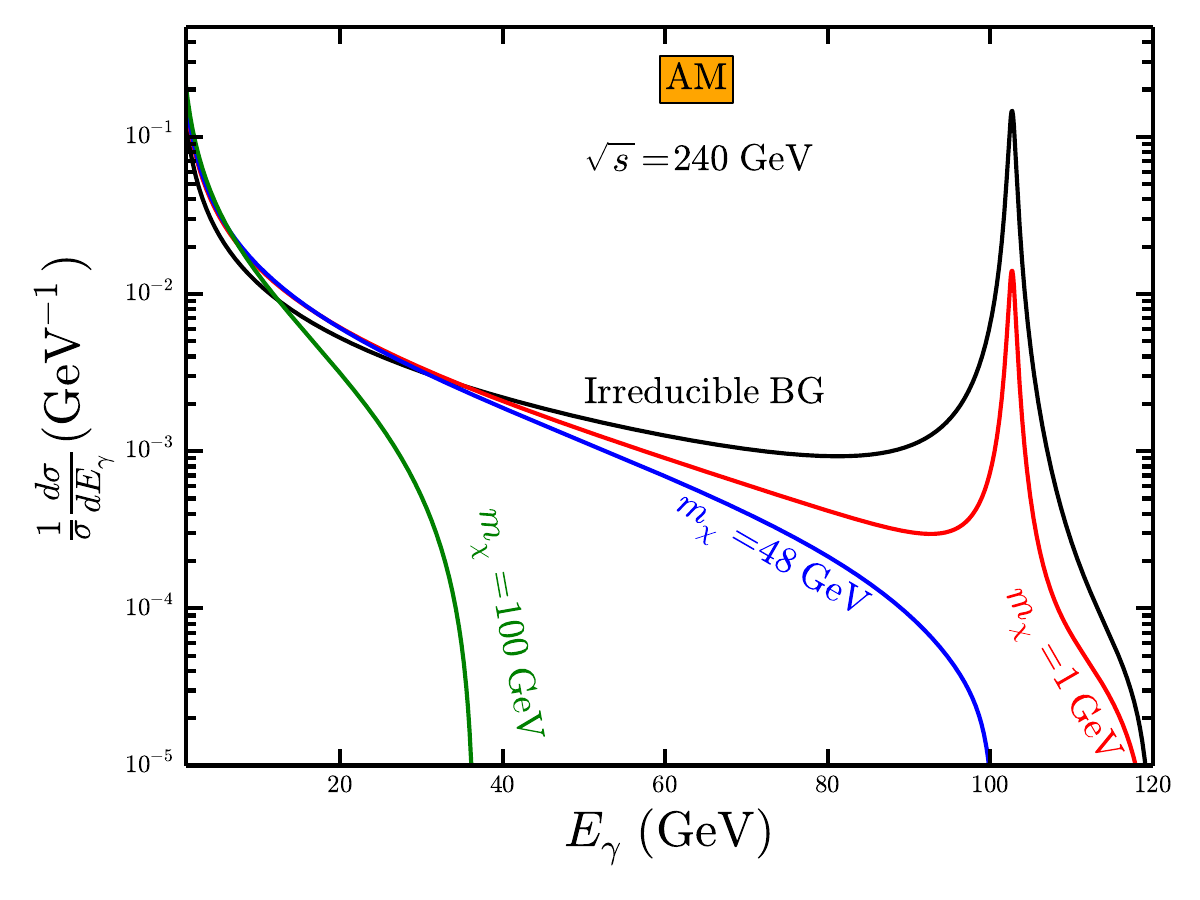}
		\caption{
Normalized $E_\gamma$ distribution  in the CEPC $H$-mode ($\sqrt{s}=240$ GeV) with detector cuts $E_\gamma>0.1$ GeV and $|\cos\theta_\gamma|<0.99$, for the irreducible background $e^+e^-\to\nu\bar\nu\gamma$ in SM and for dark states production through mass-dimension 5  MDM (left) and mass-dimension 6  AM (right).
We consider three different masses in each case with $m_\chi=$ 1 GeV, 48 GeV, and 100 GeV.}
		\label{fig:cepcdis}
	\end{centering}
\end{figure*}

 {Figure \ref{fig:cepcdis} shows the normalized energy distribution of final visible photon $E_\gamma$ 
 in the CEPC $H$-mode ($\sqrt{s}=240$ GeV) with detector cuts $E_\gamma>0.1$ GeV and $|\cos\theta_\gamma|<0.99$, for the irreducible background $e^+e^-\to\nu\bar\nu\gamma$ in SM and for dark states production through mass-dimension 5  MDM  and mass-dimension 6  AM, respectively. It can be seen that the irreducible background exhibits a resonance peak in the monophoton energy spectrum which is centered at the photon energy $E_\gamma^Z=\frac{s-M_Z^2}{2\sqrt{s}}$ with a full-width-at-half-maximum
 as $\Gamma_\gamma^Z=M_Z\Gamma_Z/\sqrt{s}$ due to the SM $Z$ boson.  We will refer to this resonance in the monophoton energy spectrum as the ``$Z$ resonance" hereafter.
 To suppress the irreducible background contribution, we will veto the events within $5\Gamma_\gamma^Z$ at the ``$Z$ resonance" in the monophoton energy spectrum (hereafter the ``$Z$ resonance veto cut").
 Though there also a  ``$Z$ resonance" in the light dark states production since the dark states can be produced via the their coupling with $Z$ boson, the ``$Z$ resonance veto cut" can also improve the ratio of signal to background.
 The final state photon in associated with dark states production can have a maximum energy as $m_\chi$, which is given by $E_\chi^m\equiv{(s-4m_\chi^2)}/{(2\sqrt{s})}$.  
 Thus, to suppress the contributions from SM, we  apply the following detector cuts at CEPC:
 \begin{itemize}
 	\item[(1)]  $E_\gamma>0.1$ GeV and $|\cos\theta_\gamma|<|\cos\theta_b|=0.99$,
 	\item[(2)]  $E_\gamma >E_\gamma^m(\theta_\gamma)= {\sqrt{s}}{(1+{\sin\theta_\gamma}/{\sin\theta_b})^{-1}}$,
 	\item[(3)] $E_\gamma < E_\chi^m={(s-4m_\chi^2)}/{(2\sqrt{s})}$,
 	\item[(4)] veto $E_\gamma\in5\Gamma_\gamma^Z$.
 \end{itemize}
  }

\begin{figure*}[htbp]
		\begin{centering}
			\includegraphics[width=0.45\columnwidth]{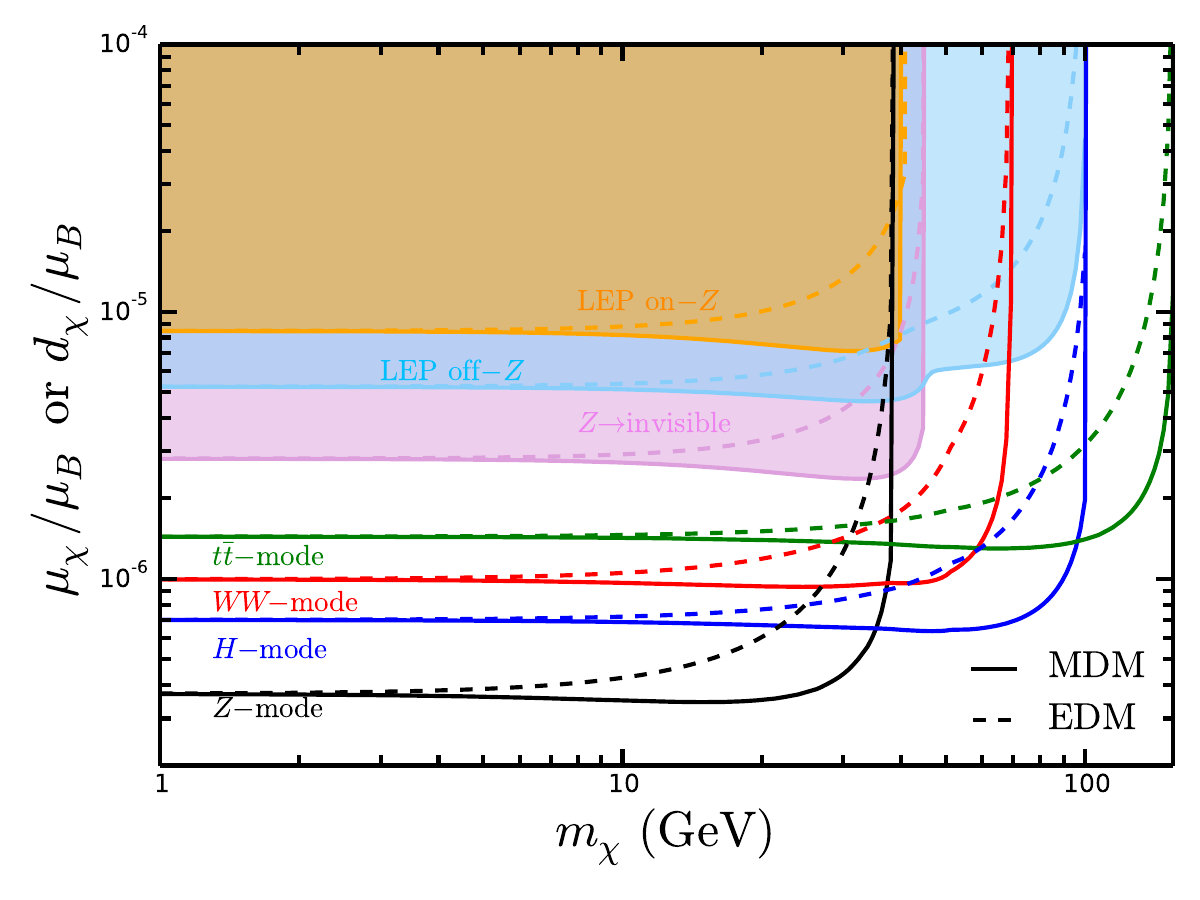}
			\includegraphics[width=0.45\columnwidth]{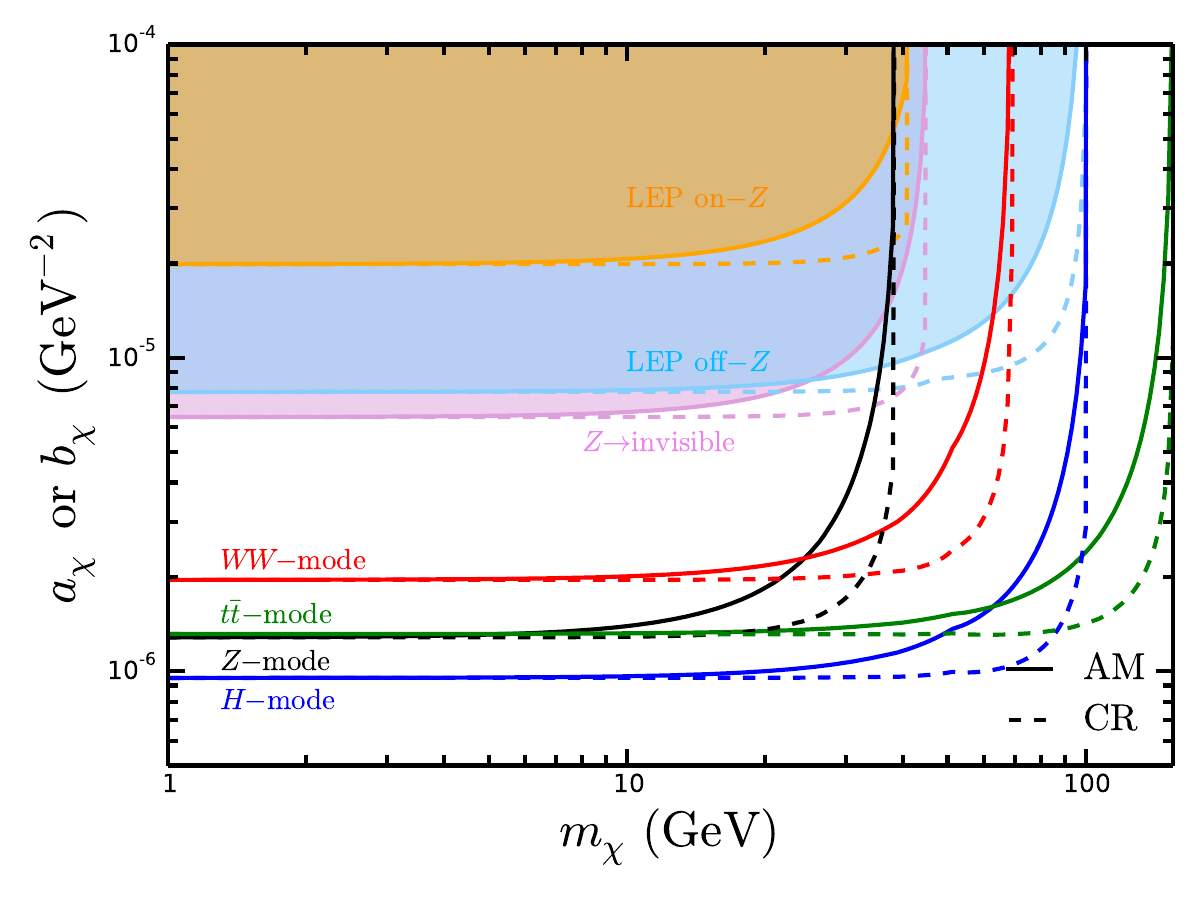}
			\caption{
				The expected 95\% C.L. upper limits on the electromagnetic form factors			
				for mass-dimension 5 operators (left) through MDM (solid) and EDM (dashed) , and for mass-dimension 6 operators (right) through AM (solid) and CR interaction (dashed)  at CEPC in the $Z$-mode with 100 ab$^{-1}$ luminosity (black), in the $H$-mode with 20 ab$^{-1}$ luminosity (blue), and in $WW$-mode with 6 ab$^{-1}$ luminosity (red), and in $t\bar t$-mode with 1 ab$^{-1}$ luminosity (green), respectively.
				The constraints at LEP from monophoton searches using the on  \cite{L3:1998uub} (orange) and off \cite{L3:2003yon} (skyblue) $Z$-pole data and the measurement of invisible decay width of the $Z$ boson \cite{ALEPH:2005ab}  (purple) are also shown with shaded regions. }
			\label{fig:cepc}
		\end{centering}
	\end{figure*}

	We use the simple criteria $S^2/B=2.71$ to study the 95\% C.L. upper bounds on the couplings at CEPC,
	which are shown in  Figure \ref{fig:cepc}. 
	Here we compute the limits based on $20\ \mathrm{ab}^{-1}$ data in the $H$-mode, $6\ \mathrm{ab}^{-1}$ data in
	the $WW$-mode, $100\ \mathrm{ab}^{-1}$ data in the $Z$-mode, and $1\ \mathrm{ab}^{-1}$ data in the $t\bar t$-mode.
	The $Z$-mode has the best sensitivity 
	for   mass-dimension 5 operators MDM with $m_\chi\lesssim$ 35 GeV, and EDM with $m_\chi\lesssim$ 25 GeV, which can probe the couplings down to about $3.7\times 10^{-7}\, \mu_B$.
	The $H$-mode has the best sensitivity 
	for MDM with 35 GeV $\lesssim m_\chi\lesssim$ {98} GeV, EDM with 25 GeV $\lesssim m_\chi\lesssim$ {79} GeV, AM with $ m_\chi\lesssim$ {63} GeV,
	and CR with $ m_\chi\lesssim$ {89} GeV, and the corresponding couplings can be probed  down to about $6.4\times 10^{-7}\, \mu_B$, $1.1\times 10^{-6}\, \mu_B$, $1.3\times 10^{-6}$ GeV$^{-2}$ and $9.8\times 10^{-7}$ GeV$^{-2}$ respectively,  
	for the case where $m_\chi\sim 50$ GeV by the $H$-mode running of CEPC.
	{Although the luminosity of $t\bar t$-mode is only one percent of that of  $Z$-mode, 
	the upper limits from $t\bar t$-mode are still comparable with that of 
	$Z$-mode for light dark states with mass-dimension 6 operators, due to the fact that the production cross sections for $\chi$ are larger and the SM irreducible background is smaller in the $t\bar t$-mode than the $Z$-mode.
	The $t\bar t$-mode has the best sensitivity for heavy  dark states $\chi$.
	With $m_\chi\sim 100$ GeV, 
	$\mu_\chi\sim 1.4\times 10^{-6}\, \mu_B$,  $d_\chi\sim 3.2\times 10^{-6}\, \mu_B$, $a_\chi\sim 2.4\times 10^{-6}$ GeV$^{-2}$, and $b_\chi\sim 1.4\times 10^{-6}$ GeV$^{-2}$ can be probed by the $t\bar t$-mode running of CEPC.}

	\section{Discussion and conclusions}
	\label{sec:con}

\begin{figure*}[htbp]
		\begin{centering}		
			\includegraphics[width=0.48\columnwidth]{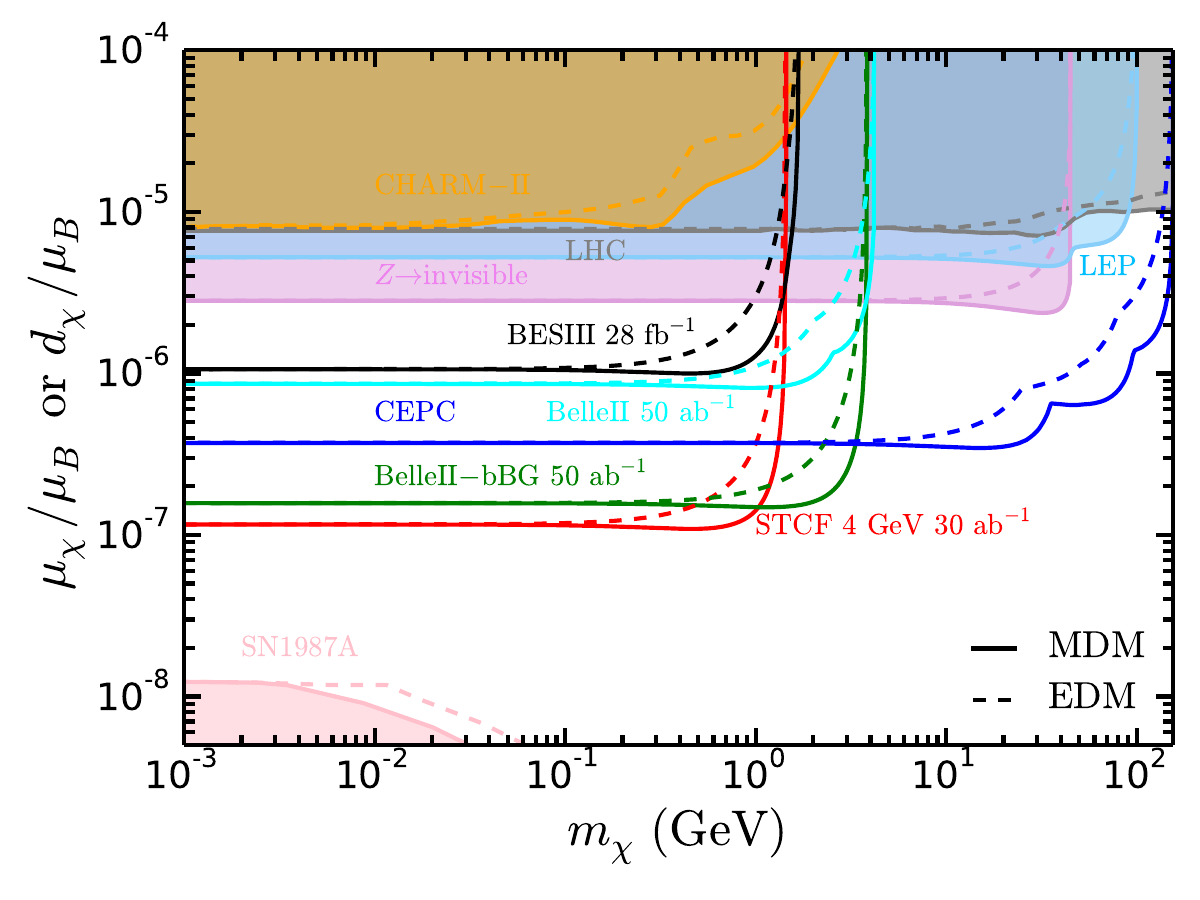}
			\includegraphics[width=0.48\columnwidth]{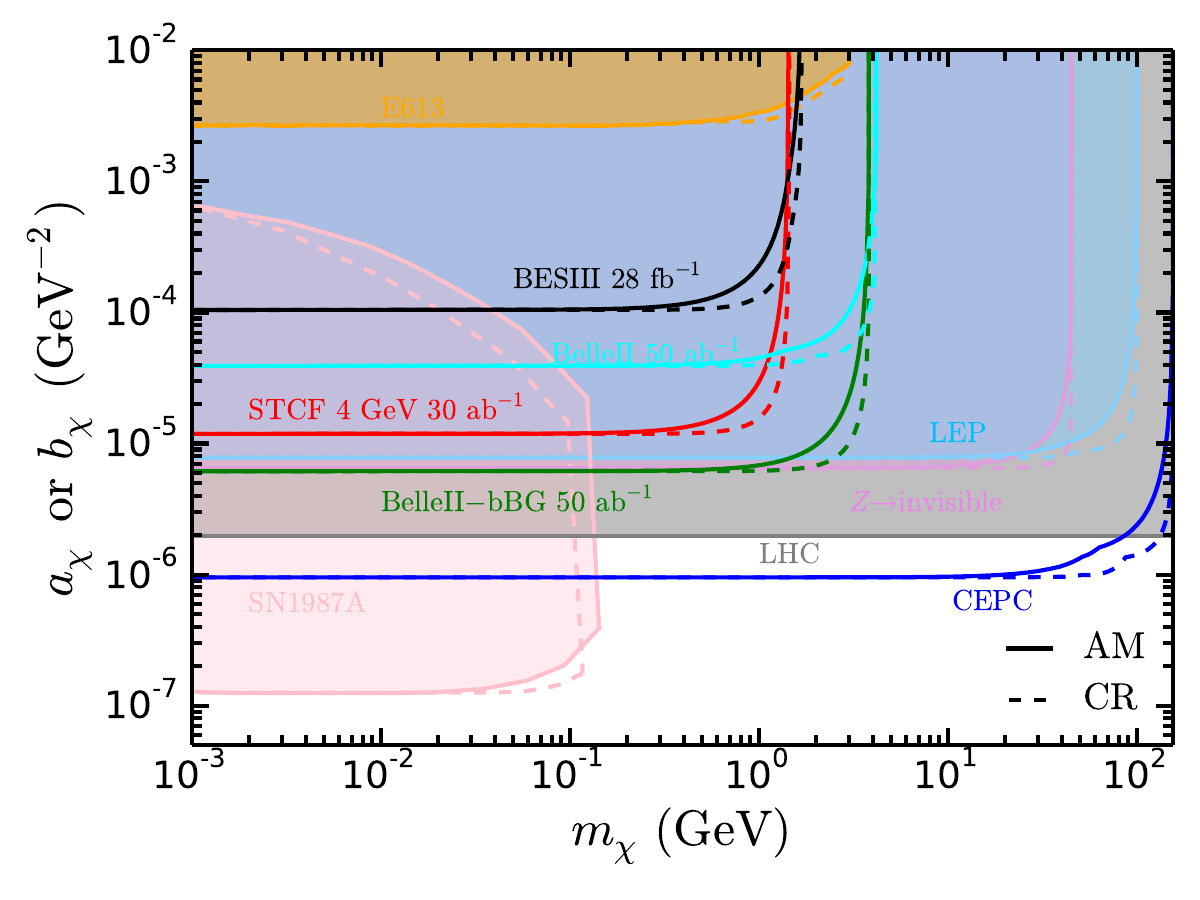}
			\caption{
				The expected 95\% C.L. exclusion limits on the electromagnetic form factors at electron colliders, including BESIII, STCF, Belle II and CEPC
				for mass-dimension 5 operators (left) through MDM (solid) and EDM (dashed) , and for mass-dimension 6 operators (right) through AM (solid) and CR interaction (dashed).
				The Belle II limits (cyan lines) combine the
				low-mass and high-mass limits in Fig. \ref{fig:belle2}, where both the bBG and
				the gBG are considered. 
				The other  Belle II limits (green lines) are obtained with the {\it ``bBG"} cut where the gBG is omitted.
				The limits from BESIII (black lines) are obtained with about 28 fb$^{-1}$ integrated luminosity collected at the various CM energies
				from 2.125 GeV to 4.95 GeV during 2012-2021. 
				The STCF limits
				(red lines) are obtained with $\sqrt{s}=$ 4 GeV and 30 ab$^{-1}$. 
				The CEPC limits (blue lines) combine the best limits in $Z$-mode, $H$-mode, and $t\bar t$-mode in Fig. \ref{fig:cepc}.
				The gBG is not considered at BESIII, STCF and CEPC.
				The landscape of current leading constraints are also shown  with shaded regions, exploiting from terrestrial experiments, 
				such as proton-beam experiments CHARM-II  or E613 \cite{Chu:2020ysb}, monophoton searches and $Z$-boson invisible decay at LEP  and monojet searches at LHC \cite{Arina:2020mxo}, and astrophysics supernovae SN 1987A \cite{Chu:2019rok}. 
			}
			\label{fig:total}
		\end{centering}
	\end{figure*}

	The landscape of current excluded parameter space in the plane of dark states mass and coupling to the photon  with mass-dimension 5 (left panel) operators through MDM (solid) and EDM (dashed) and with mass-dimension 6 (right panel) operators through AM (solid) and CR interaction (dashed) are shown in Fig. \ref{fig:total} by shaded regions, obtaining from terrestrial experiments, 
	such as proton-beam experiments CHARM-II  or E613 \cite{Chu:2020ysb}, monophoton searches and $Z$-boson invisible decay at LEP  and monojet searches at LHC \cite{Arina:2020mxo}, and astrophysics supernovae SN 1987A \cite{Chu:2019rok}. 
	It should be noted that we only show several more competitive constraints in Fig. \ref{fig:total},
	the more complete results can be found in Refs. \cite{Chu:2018qrm, Chu:2019rok, Chu:2020ysb,Kling:2022ykt}.
	The 95\% C.L. constraints on the dark states with electromagnetic form factors
	derived above  from the electron colliders, BESIII, STCF, Belle-II, and CEPC are also plotted with lines in Fig. \ref{fig:total}.
	The Belle II limits (cyan  lines) combine the low-mass and high-mass limits in Fig. \ref{fig:belle2}, where both the bBG and
	the gBG are considered. 
	{To investigate the possible potential of Belle II, and compare the sensitivity on the dark states-photon couplings with other electron colliders whose detailed simulations on gBG are not available, we also present the  limits at  Belle II (green curves) with gBG
		omitted.
		It's noted that the actual limits from BESIII, STCF, and CEPC should be weaker when gBG
		is taken into account.  }

	For the mass-dimension 5 operators, $Z$ boson invisible decay is  most sensitive with $m_\chi \lesssim 45$ (40) GeV through MDM (EDM) among the current constraints from the terrestrial experiments mentioned above.
	Monophoton search at the LEP  is  currently the strongest  constraint in the range of  45 GeV $\lesssim m_\chi \lesssim 100$ GeV through MDM.
	BESIII can probe new parameter space that is previously unconstrained
	by other experiments for mass $\lesssim$ 1 GeV, with 28 fb$^{-1}$ data collected during 2012-2021. 
	BESIII with the omission of the gBG (28 fb$^{-1}$) only leads to
	a slightly weaker limit than Belle II (50 ab$^{-1}$) with gBG included
	for  $m_\chi \lesssim 1$ GeV. 
 Although the STCF luminosity  (30 ab$^{-1}$) is lower than
	Belle II (50 ab$^{-1}$), STCF has better sensitivity in probing the low-mass region ($m_\chi \lesssim 1$ GeV) than Belle II,
	{if we assume that the gaps in the detector can be significantly suppressed
	in the future experiments, for instance, with a new
	subdetector that can detect the particles emitting from the
	gaps in ECL. }
	This is because 
	STCF is operated at a lower colliding energy ($\sqrt s=4$ GeV) 
	where the monophoton cross section in SM{, mainly coming from irreducible neutrino backgrounds since the reducible QED backgrounds can be removed by the bBG cut,} is smaller than Belle II ($\sqrt s=10.58$ GeV), and $\chi\bar\chi\gamma$ production rate is not very dependent on the CM energy for mass-dimension 5 operators.
	{It is noted that the monophoton production rates  from the reducible QED backgrounds, such as radiative Bhabha scattering $e^+e^-\to e^+e^-\gamma$, will  grow with lower  CM energies \cite{Zhang:2019wnz}, thus potentially reducing the low energy advantage of STCF over Belle II.
	}
		The about five times of magnitude difference in
	sensitivity between the two Belle II limits, the cyan curve
	and the green curve in Fig. \ref{fig:total}, shows that the control
	on gBG is very important in probing the electromagnetic form factors via mass-dimension 5 operators. 
	When the background due to the gaps in the detectors is neglected, 
	the future CEPC can give leading constraints than other electron colliders when $m_\chi\gtrsim 4$ GeV, which can probe the coupling down to $3.7\times 10^{-7}\, \mu_B$.
	While with about 100 ab$^{-1}$ luminosity running at 91.2 GeV, the bounds on the light dark states from CEPC are still weaker
	than the ones from STCF and Belle-II with gBG
	omitted. It implies that the low-energy electron colliders can secure a place in the future to probe low-mass
	light dark states with electromagnetic form factors via mass-dimension 5 operators{, if the main reducible QED gBG can be significant suppressed, since  there is significant uncertainty in
		understanding the reach of BESIII/STCF given that the main background
		rates are not known.}.

	With regard to the  mass-dimension 6 operators, the bounds from the mono-jet search at LHC constrain better than other current experimental sensitivity in the plotted region in Fig.\ref{fig:total}, expect for the light dark states $\chi$ with $m_\chi \lesssim 10$ MeV, which are constrained dominantly by astrophysical bound from SN1987A. The upper limits from low-energy electron colliders, such as BESIII, STCF and Belle II (except Belle II with gBG omitted), are all excluded by the monophoton search at LEP.
	This is because that, for mass-dimension 6 operators, the production 
	rates of light dark states $\chi$ are even more sensitive to the
	CM energy, suggesting that it is unlikely for low-energy experiments to play any role in the foreseeable future. 
	The high-energy colliders, such as CEPC, can probe a vast region of the parameter
	space that is previously unexplored, for the light dark states with electromagnetic form factors via mass-dimension 6 operators through AM (CR) in the mass region from
	20 MeV to 90 (140) GeV. Compared to current LHC bounds, the improvement on upper limits of couplings is about two times of magnitude for the mass less than 10 GeV.

	\acknowledgments
	This work was supported in part by the National Natural Science Foundation of China (Grants No.12105327, No. 11805001), the Natural Science Foundation of Anhui Province (No.2208085MA10), and the Key Research Foundation of Education Ministry of Anhui Province of China (No.KJ2021A0061).


\end{document}